\documentclass[journal]{IEEEtran}
\usepackage{pifont}
\usepackage{cite}

\usepackage[dvips]{graphicx}
 \graphicspath{{/Figure/}}
\DeclareGraphicsExtensions{.eps}

\usepackage{array}
\usepackage{multirow}
\usepackage{color}
\usepackage{longtable,balance}
\usepackage{graphicx}
\usepackage{supertabular}
\usepackage[usestackEOL]{stackengine}
\usepackage{epic}
\usepackage{tabularx}
\usepackage{subfigure}
\usepackage{booktabs}
\usepackage{amsmath}
\usepackage{amssymb}
\usepackage{amsopn}
\usepackage{enumerate}
\usepackage{booktabs}
\usepackage{eso-pic}

\begin{document}
\title{\huge{vDLT: A Service-Oriented Blockchain System with Virtualization and Decoupled Management/Control and Execution}}  

\author{Version 0.1 \\ \vspace{0.5cm} \IEEEauthorblockN{F. Richard Yu,}~\IEEEmembership{IEEE Fellow} \\
\IEEEauthorblockA{vDLT Laboratory, Ottawa, ON, Canada}\\

}
\maketitle

\begin{abstract}
A wide range of services and applications can be improved and/or solved by using distributed ledger technology (DLT). These services and applications have widely varying quality of service (QoS) requirements. However, most existing DLT systems do not distinguish different QoS requirements, resulting in significant performance issues such as poor scalability and high cost. In this work, we present vDLT -- a service-oriented blockchain system with virtualization and decoupled management/control and execution. In vDLT, services and applications are classified into different classes according to their QoS requirements, including confirmation latency, throughput, cost, security, privacy, etc. This is a paradigm shift from the existing ``blockchain-oriented" DLT systems to next generation ``service-oriented" DLT systems. Different QoS requirements are fulfilled by advanced schemes inspired by the development of the traditional Internet, including classification, queuing, virtualization, resource allocation and orchestration, and hierarchical architecture. In addition, management/control and execution of smart contracts are decoupled to support QoS provisioning, improve decentralization, and facilitate evolution in vDLT. With virtualization, different virtual DLT systems with widely varying characteristics can be dynamically created and operated to accommodate different services and applications. 
\end{abstract}



\section{Introduction}
Recently, \emph{distributed ledger technology} (DLT) (e.g., blockchain) has attracted great attentions from both industry and academia \cite{Bec18}. Similar to TCP/IP (transmission control protocol/Internet protocol), which laid the groundwork for the development of the Internet, DLT has great potential to create new foundations for our socio-economic systems by efficiently establishing trust among people and machines, reducing cost, and increasing utilization of resources \cite{IL17}. With the rise of DLT, socio-economic transactions are improving as we shift from the \emph{Internet of information} (IoI) to the \emph{Internet of value} (IoV). 

A wide range of services and applications can be improved and/or solved by using DLT. Although the first killer application of DLT is cryptocurrency (e.g., Bitcoin \cite{Bitcoin}), the underlying constructs do not have to be limited to payment transactions. The services and applications of DLT include supply chain management, identification, healthcare, music, energy, gaming, agriculture, transportation, publishing, etc. The `World Economic Forum' anticipates that 10\% of global GDP will be stored on the blockchain by 2025. The impact of DLT could be as grand as the traditional Internet revolution itself.

Nevertheless, a number of non-trivial issues in the current DLT systems prevent them from being used as a generic platform for different services and application across the globe. One notable drawback is the scalability issue. Bitcoin can process about 7 transactions per second (TPS), and Ethereum has the ability of processing about 15 TPS, which is far below the mainstream payment systems, e.g., VISA with more than 2,000 TPS capability. With even one popular application (e.g., CyptoKitties in Dec. 2017 and FCoin in July 2018), Ethereum can be severely congested with significantly increased delay and transaction fee.  

There is no silver-bullet that solves all these problems due to the \emph{Trilemma} as described by Vitalik Buterin, the founder of Ethereum: DLT systems can only at most have two of the following three properties: decentralization, scalability and security. Most of the recently developed DLT systems focus on increasing transaction throughput to improve scalability, e.g., Lightning Network \cite{LightingNetwork}, Raiden Network \cite{RaidenNetwork}, Sharding and Plasma \cite{Plasma}, Cardano \cite{Cardano}, EOS \cite{EOS}, Zilliqa \cite{Zilliqa}, etc.

Similar issues occurred in the development of the traditional Internet. In the 1990s, with more and more applications built on TCP/IP, the Internet became often congested, and the performance of some applications (e.g., video streaming) was not acceptable for massive popularity due to network congestion \cite{Jai90} \footnote{Youtube was not launched until 2005.}. With the rapid transformation of the Internet into a commercial infrastructure, demands for service quality have rapidly developed. One intuitive solution was to increase the link bandwidth (and hence the throughput) by deploying fibers and wavelength-division multiplexing (WDM). People believed that, with bandwidth so abundant, the quality of service (QoS) will be automatically delivered. This solution is very similar to ideas behind most of the recently proposed DLT systems, i.e., increasing TPS. Indeed, TPS has been regarded as one of the most important parameters in designing a DLT system.

However, the history of the traditional Internet has told us that increasing throughput alone cannot solve the congestion problem. Even worse, increasing throughput without proper QoS designs may aggravate the congestion problem \cite{Jai92}. There are several reasons for this: heterogeneous QoS requirements from different applications, dynamics of applications, dynamics of available resources, distributed networks without central coordination, etc. \cite{Jai90, Jai92}. 

In DLT systems, these situations still apply. For example, different services and applications built on DLT have widely varying QoS requirements. While instant confirmation is desirable when you are buying a cup of coffee using cryptocurrencies, confirmation latency can be tolerated when you are buying a house or conducting computation-intensive machine learning tasks. Moreover, in addition to TPS, other metrics should be considered, such as cost (e.g., transaction fee (a.k.a. gas) in Ethereum and RAM costs in EOS). While it may be ok to pay \$1 transaction fee to buy a cup of coffee, it is undesirable to pay \$1 for transferring several bits (e.g., reading temperature) in Internet of things (IoT) applications with billions of IoT devices, or \$1 for creating an account in social media applications with billions of users. Furthermore, while privacy is the main concern in some applications, others may not care about privacy.

To address these issues, we present vDLT -- a service-oriented blockchain system with virtualization and decoupled management/control and execution. The distinct features of vDLT are as follows.

\begin{itemize}
\item Unlike most  existing DLT systems that do not distinguish different services and applications, vDLT explicitly considers the QoS requirements of different services and applications. Specifically, services and applications are classified into different classes according to their QoS requirements, including confirmation latency, throughput, cost, security, privacy, etc. 
\item This is a paradigm shift from the existing ``blockchain-oriented" DLT systems to next generation ``service-oriented" DLT systems.
\item Different QoS requirements are fulfilled by advanced schemes inspired by the development of the traditional Internet, including classification, queuing, virtualization, resource allocation and orchestration, and hierarchical architecture.
 
\item Management/control (e.g., governance, smart-contract-execution nodes selection, and resource allocation) and execution of smart contracts are decoupled to support QoS provisioning, improve decentralization, and facilitate evolution in vDLT.

\item With virtualization, different virtual DLT systems with widely varying characteristics can be dynamically created and operated to accommodate different services and applications. 

\end{itemize}

This document outlines the technical design of vDLT. The rest of this document is organized as follows. The related work is presented in Section II. Section III describes the system overview of vDLT. The vDLT design details are presented in Section IV. Finally, we conclude this work in Section V.
 

\section{Related Works in Telephone Networks, the Traditional Internet, and Cellular Networks}
In this section, we briefly review telephone networks, the traditional Internet (i.e., the Internet of information) and cellular networks. From the history of telephone networks, the traditional Internet, and cellular networks, we can see that, at the beginning of the development of these systems, management/control and user traffic were usually coupled together due to easier implementation. However, as the system evolved over time, management/control is decoupled from user traffic due to many benefits described below. Table I summarizes this process.

\begin{table*}
\label{Table_Decoupling}
\centering
  \caption{Decoupling Control from User Traffic in Telephone Networks, the Traditional Internet and Cellular Networks.}\label{Decoupling.}
  \begin{tabular}{|p{2.17cm} | l | p{4.7cm} | l |}
\hline
 & Before the Decoupling & After the Decoupling  &  Benefits of the Decoupling
\\ \specialrule{1.3pt}{1pt}{1pt} 
\multirow{3}{*} {Telephone Networks} &  & \multirow{3}{*} {$\bullet$ Signaling System No. 7 (SS7)} & $\bullet$ Reduce the call setup time \\
& \multirow{1}{*} {$\bullet$ Signaling System No. 5 } & & $\bullet$ Reduce the toll fraud \\
& (SS5) & & $\bullet$ Easier to introduce new services 

\\ \hline

\multirow{3}{*} {Traditional Internet} & $\bullet$ Best-effort & $\bullet$ Network Function Virtualization (NFV) &$\bullet$ Lower operation cost \\
 & $\bullet$ InterServ & $\bullet$ Software-defined Networking (SDN) & $\bullet$ Simplify network management \\
 & $\bullet$ DiffServ &  & $\bullet$ Facilitate network evolution

\\ \hline
\multirow{3}{*} {Cellular Networks} & \multirow{3}{*} {$\bullet$ 4th Generation (4G)} &  & $\bullet$ Reduce latency of applications and services \\
& & \multirow{1}{*} {$\bullet$ Control\&User Plane Separation  (CUPS) } & $\bullet$ Increase throughput \\
& & in 5G  & $\bullet$ Independent evolution of control and user planes 

\\ \hline

\end{tabular}
\end{table*}

\subsection{Decoupling Control from User Traffic in Telephone Networks}

Commercialization of the telephone began in 1876, with instruments operated in pairs for private use between two locations. Before the 1970s, the public switched telephone network (PSTN) used in-band signaling, which is the exchange of call control information (e.g., telephone number) within the same channel that the user telephone call (traffic) itself is using. An example is dual-tone multi-frequency signaling (DTMF) used in Signaling System No. 5 (SS5). In-band signaling is insecure because it exposes control signals, protocols and management systems to end users. In addition, it is inflexible for operators to introduce new services.

Out-of-band signaling is transmitted over a dedicated channel separated from that used for the telephone call. Out-of-band signaling has been used since SS6 was introduced in the 1970s, and also in SS7 \cite{UGM94} in 1980, which became the standard for signaling among exchanges ever since. By decoupling management/control from user traffic, out-of-band signaling can significantly  reduce the call setup time and toll fraud. In addition, with this decoupling, it is much easier for the operators to introduce new services, including 800\# portability, wireless roaming, caller ID and other CLASS (Custom Local Area Signaling Services) services \cite{BDO92}.

\subsection{Quality of Service Provisioning in the Traditional Internet}
The circuit switching technology of telephone networks was woefully inadequate for supporting data communications. TCP/IP was proposed in the 1970s as a suite of communication protocols used to interconnect network devices on the Internet. Only best-effort service was provided in the original design of the traditional Internet, where management/control and traffic are coupled together. With the rapid transformation of the Internet into a critical infrastructure with a wide range of applications, demands for QoS had rapidly developed. Several service classes were demanded. For example, one service class can provide predictable Internet services with interactive applications (e.g., Web). Another service class can provide low-delay and low-jitter services (e.g., Internet telephony and videoconferencing). Best-effort service will remain for those applications that just need connectivity.

Whether service classification and QoS mechanisms are even needed was a hotly debated issue in the community. One opinion was that increasing link capacity via fibers and wavelength-devision multiplexing (WDM) will make bandwidth so abundant, and QoS will be automatically delivered. The other opinion was that, no matter how much bandwidth the network can provide, new applications will be invented to consume it, and efficient QoS mechanisms will still be needed. It was shown that increasing link bandwidth, memory sizes, processor speeds cannot effectively address the QoS issues. Even worse, increasing these resources without proper QoS designs may aggravate the congest problem \cite{Jai92}. There are several reasons for this: heterogeneous QoS requirements from different applications, dynamics of applications, dynamics of available resources, distributed networks without central coordination, etc. \cite{Jai90, Jai92}.

To address the QoS issues, several service models and mechanisms have been proposed. Notably among these are the \emph{integrated services} (IntServ) model \cite{IntServ}, the \emph{differentiated services} (DiffServ) model \cite{DiffServ}, and \emph{network function virtualization} (NFV) \cite{CO16,WRH15} and \emph{software-defined networking} (SDN) \cite{NMN14, WRH15}. In the IntServ model, applications ask the network for an explicit resource reservation per flow, which is defined by source and destination IP addresses and ports. By reserving resources in the network for each flow, applications have resources guarantees and predictable behaviors. Although IntServ model can provide hard QoS guarantees, the poor scalability issue makes it difficult to deploy IntServ model in large-scale networks. By contrast, Diffserv model is a soft QoS model, which is based on service classes and per hop behaviors associated to each class. DiffServ allows to classify packets into different treatment categories, each of which will receive  different per hop behaviors at each hop from the source to the destination. Although DiffServ model scales well in large-scale networks, it cannot provide hard QoS guarantees.

\subsection{Virtualization and Decoupling Control from User Traffic in the Traditional Internet}

\begin{figure}[tp]
  \centering
	\vspace{-14.5cm}
	\hspace{5cm}
  \includegraphics[width=0.83\textwidth]{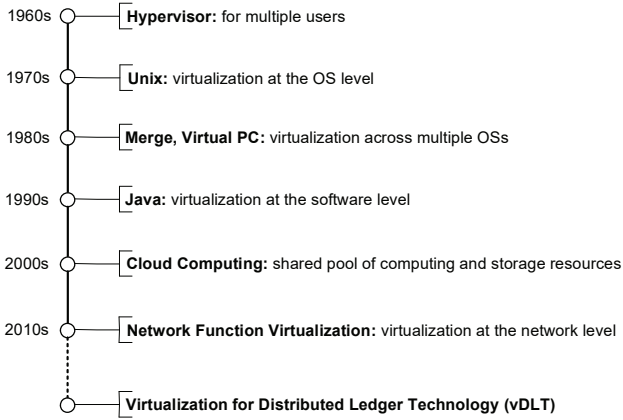}
	\vspace{-0.3cm}
		\caption{A brief journey of virtualization technologies in the IT world.}
  \label{fig_virtu_history}
\end{figure}

\emph{Virtualization} has been revolutionizing the IT world, including the recent advances of cloud computing \cite{Sal12}, edge computing \cite{MCD18}, and network function virtulization (NFV) \cite{HGJ15}. Figure \ref{fig_virtu_history} shows a brief journey of virtualization in the IT world. Essentially, virtualization refers to technologies designed to provide  abstraction of underlying resources (e.g., hardware, compute, storage, network, etc.). 

\begin{figure*}[tp]
  \centering
	\vspace{-11cm}
	\hspace{0cm}
  \includegraphics[width=0.9\textwidth]{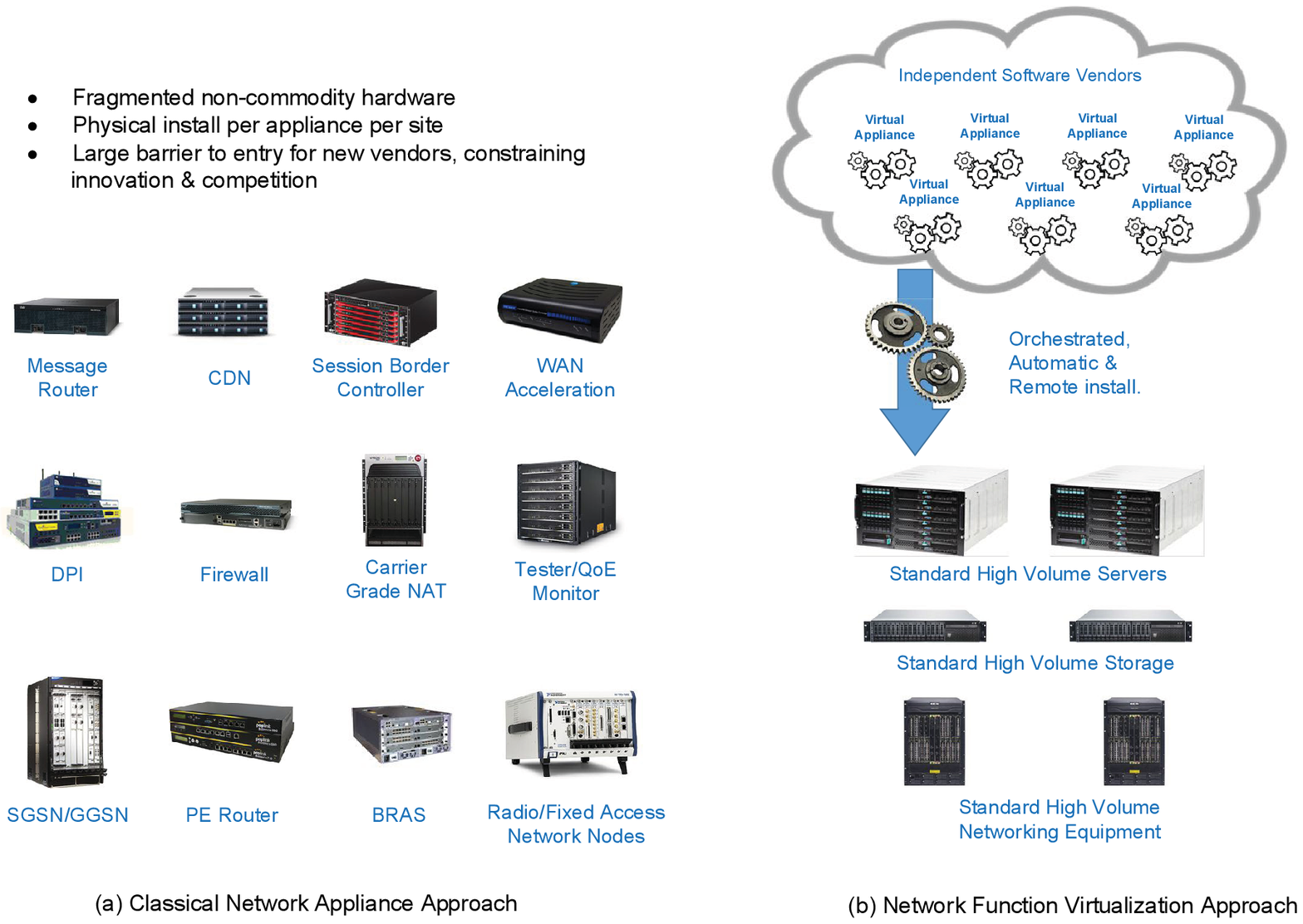}
	\vspace{0cm}
		\caption{Comparison between (a) classical network appliance approach and (b) network function virtualization (NFV) approach.}
  \label{fig_nfv}
\end{figure*}

With the tremendous growth in the Internet traffic and services, it is natural to extend the success of virtualization from computing and storage to networks. Recently, network virtualization has been actively used in Internet research testbeds, such as G-Lab \cite{G-Lab} and 4WARD \cite{achemlal2009virtualisation}. It aims to overcome the resistance of the current Internet to fundamental architecture changes. Network virtualization has been considered as one of the most promising technologies for the future Internet \cite{LY15}. Particularly, the NFV concept was presented by a group of network service providers in 2012. These service providers wanted to simplify and speed up the process of adding new network functions or applications. The European Telecommunications Standards Institute (ETSI) Industry Specification Group for Network Functions Virtualization proceeded to spearhead NFV development and standards \cite{HGJ15}.

In traditional networks, network services are run on proprietary, dedicated hardware. With NFV, functions like routing, load balancing and firewalls are packaged as virtual machines (VMs) on commodity hardware. Individual virtual network functions (VNFs), are an essential component of NFV architecture. Because NFV architecture virtualizes network functions and eliminates specific hardware, network managers can add, move or change network functions at the server level in a simplified provisioning process. Figure \ref{fig_nfv} show the comparison between the traditional network appliance approach and the NFV approach.

Instead of considering all the functions of networking, SDN focuses on two main  functions, control and traffic forwarding, in the design. Specifically, the control plane and traffic forwarding plane are decoupled in SDN. Compared to traditional networking paradigms, SDN makes it easier to introduce new abstractions in networking, lowering operation costs, simplifying network management, and facilitating network evolution \cite{NMN14, WRH15}.

\subsection{Virtualization and Decoupling Control from User Traffic in Cellular Networks}
Virtualization has been widely adopted in cellular networks, as evidenced by the booming business of mobile virtual network operators (MVNOs), such as Tracfone, Virgin Mobile, and  Boost Mobile. A MVNO is a wireless communications services provider that does not own the wireless network infrastructure over which it provides services to its customers. Virtualization technologies enable MVNOs to launch new services faster to accommodate different QoS requirements of end users with lower capital expenses and operation expenses compared to their infrastructure counterparts \cite{LY15}.

In wireless cellular networks, decoupling management/control from user traffic has been always a trend. Recently, control and user plane separation (CUPS) has been adopted in the 5th generation (5G) cellular networks \cite{CUPS}. CUPS enables flexible network deployment and operation, by distributed or centralized deployment and the independent scaling between control plane and user plane functions. With CUPS, latency of applications and services can be reduced, e.g. by selecting user plane nodes that are closer to the radio access network (RAN) or more appropriate for the intended user equipment (UE) usage type without increasing the number of control plane nodes. Data traffic throughput can be increased, by enabling to add user plane nodes without changing the number of nodes in the network. By locating and scaling the control and user plane resources independently, CUPS can also facilitate independent evolution of the control plane and user plane functions. In addition, CUPS enables SDN to deliver user plane data more efficiently.

\section{vDLT System Overview}

In this section, we present an overview of the proposed vDLT system, including classification, architecture, and consensus mechanisms. Detailed description will be presented in the next section.

\subsection{Services and Applications Classification}

While user traffic (e.g., voice, video, and data) is the main concern in the traditional Internet of information, smart contracts are the main use case of DLT systems. Smart contracts are lines of code that are stored on a DLT system and automatically execute when predetermined terms and conditions are met. Different services and applications built on DLT have widely varying QoS requirements. In vDLT, services and applications are classified into different classes according to their QoS requirements, including confirmation latency, throughput, cost, security, privacy, etc. When a transaction is generated by a service or application from a node, a class for this transaction is assigned by this node. Then, this transaction will be treated differently according to the class in the vDLT system. In addition, the node generating the transaction may be a malicious node or has low trust value. Therefore, the vDLT system can ignore the class value, and assign a different class value for the transaction.

\subsection{Decoupling Management/Control from Execution}

\begin{figure}[tp]
  \centering
	\vspace{-7.6cm}
	\includegraphics[width=1.3\textwidth]{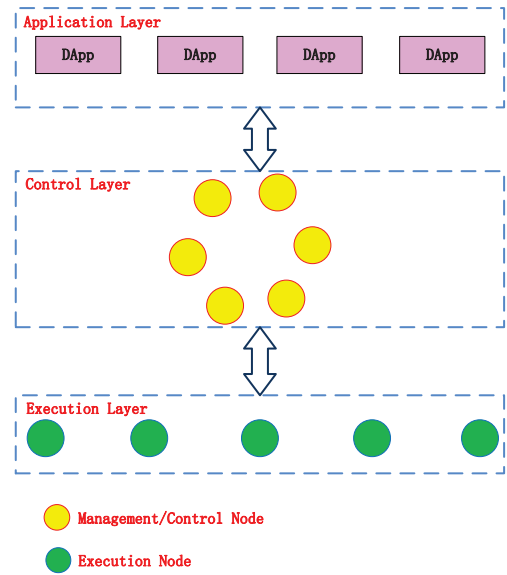}
	\vspace{-0.7cm}
		\caption{Decoupling management/control from execution in vDLT.}
  \label{fig_decoupling}
\end{figure}

The classified transaction will be sent to a group of management/control nodes, who are responsible for the management/control functions, including the prioritization of transactions, resource allocation, and the decisions on which nodes should execute the smart contract in the transaction. This group of management/control nodes conduct a blockchain consensus mechanism, and manage the health of the participants. After the consensus is reached, the transaction is sent to a group of execution nodes, who are responsible for the execution of smart contacts. Similarly, this group of execution nodes conduct a blockchain consensus mechanism to produce user transaction blocks. Please note, unlike some other blockchain systems (e.g., EOS and Dash), the management/control nodes do not produce user transaction blocks, which will be produced by the execution nodes in vDLT. This will improve decentralization of vDLT, and address the centralization issues criticized by the community. Figure \ref{fig_decoupling} shows the architecture of decoupling management/control from execution in vDLT.

\subsection{Virtualization}
From Figure \ref{fig_virtu_history}, we can see that virtualization has been playing an important role to abstract the underlying resources, so that people can focus on the things they care the most. Therefore, we believe that virtualization will be naturally the next step for DLT to address the current issues of DLT systems.

\begin{figure}[tp]
  \centering
	\vspace{-9cm}
	\includegraphics[width=1.35\textwidth]{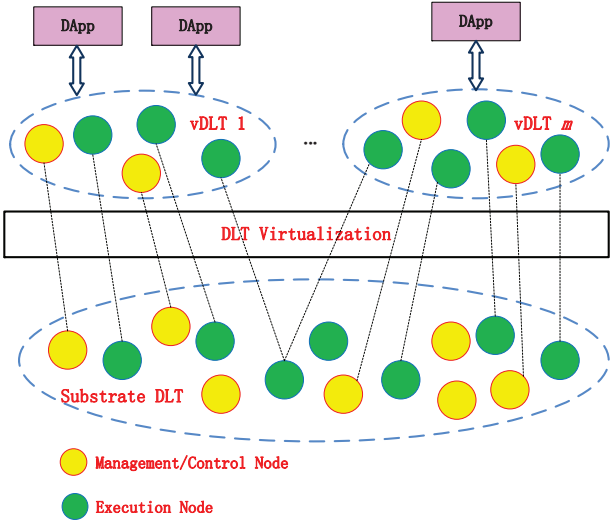}
	\vspace{-0.5cm}
		\caption{Virtual DLT systems mapped onto one substrate DLT systems.}
  \label{fig_VT}
\end{figure}

With virtualization, the underlying system resources (e.g., hardware, compute, storage, network, etc.) are abstracted. A virtual DLT system is a combination of system resources on top of a substrate DLT system, as shown in Figure \ref{fig_VT}. To accommodate different QoS requirements of different services and applications, multiple virtual DLT systems with widely varying characteristics can be created and co-hosted on the same substrate DLT system. 

Some existing DLT systems, e.g., EOS \cite{EOS}, have made fine steps in this direction. For example, a user of EOS can ``stake" his/her EOS tokens to reserve the resources (RAM, CPU, bandwidth, and storage) in the blockchain and is granted access to the reserved resources based on the amount of the staked tokens. Compared with EOS, the abstraction introduced by the virtualization mechanism allows vDLT to manage the resources in  the system in a more flexible and dynamic way.

Furthermore, in most existing DLT systems, infrastructure and service are coupled together, which makes it difficult to accommodate different QoS requirements, as evidenced by undesirable network congestion, long confirmation latency, and high cost in some DLT applications. Specifically, for a DApp, it is difficult to control confirmation latency and cost for its users due to the inflexibility of existing DLT systems.

\begin{figure}[tp]
  \centering
	\vspace{-13.5cm}
	\includegraphics[width=1.5\textwidth]{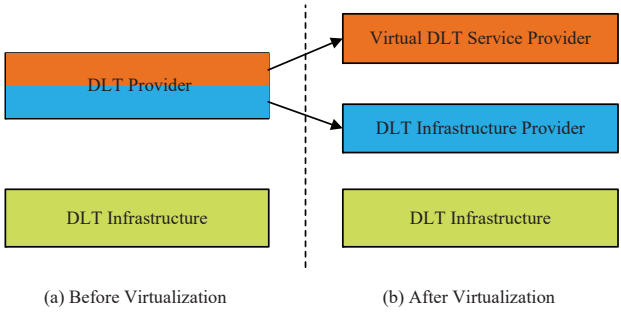}
	\vspace{-0.5cm}
		\caption{DLT business models (a) before virtualization and (b) after virtualization.}
  \label{fig_BM}
\end{figure}

With virtualization, the role of a DLT provider can be decoupled into two specialized roles, virtual DLT service provider (vDSP) and DLT infrastructure provider (DInP), as shown in Figure \ref{fig_BM}. Virtualization technologies enable vDSPs to launch new services faster to accommodate different QoS requirements of end users with lower capital expenses and operation expenses compared to their infrastructure counterparts, as seen from the great success of MVNOs.

\subsection{Consensus}
Management/Control nodes operate as part of the delegated proof-of-stake (DPoS) consensus mechanism. Under DPoS, community members will vote on delegates to represent them on the system, and these delegates are charged with management/control functions. Unlike the plurality/majority voting systems used by most DPoS systems, quadratic voting \cite{LW17} with token lock \cite{QV16} is used in vDLT.  Quadratic voting provides a better way to make collective decisions that avoids the tyranny of the majority. For the consensus among management/control nodes,  improved practical byzantine fault tolerance (PBFT) protocol is used with EC-Schnorr multi-signature \cite{STV16,KJG16}.


\section{vDLT Design Details}
In this section, we describe the design details of vDLT.

\subsection{Classification and Queuing of Transactions}

\begin{figure}[tp]
  \centering
	\vspace{-12cm}
	\includegraphics[width=1.3\textwidth]{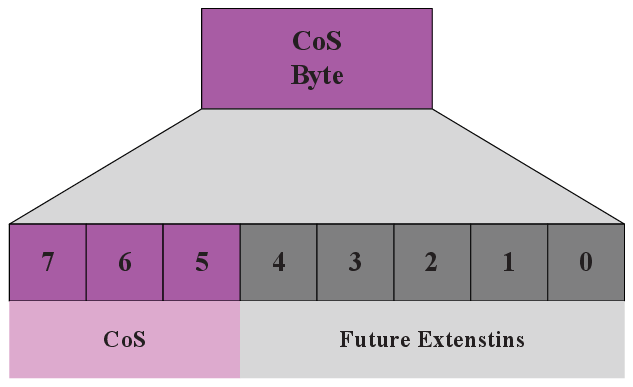}
	\vspace{-0.7cm}
		\caption{Class of service (CoS) byte in vDLT.}
  \label{fig_CoS}
\end{figure}

A class of service (CoS) byte is defined for each transaction in vDLT, as shown in Figure \ref{fig_CoS} and Table II. Due to the fact that it is difficult to predict future services and applications in DLT systems, we focus on the existing representative services and applications in the current design. Specifically, the three most significant bits of the CoS byte are used to indicate different classes. The rest bits in the CoS byte will be used for future extensions, which will be compatible with the current design. In this version, we classify services and applications into the following 8 classes: `\emph{fast confirmation}', `\emph{computation-intensive}', `\emph{storage-intensive}', `\emph{low cost}', `\emph{management/control}', `\emph{Private}', `\emph{best effort}', and `\emph{scavenger}' applications, which are described as follows.

\begin{itemize}
\item `\emph{Fast confirmation}' applications require instant confirmation for the transaction. Confirmation latency is the main concern of these applications, e.g., finance and retail applications.
\item `\emph{Computation-intensive}' applications require extensive computational resources. Decentralized machine learning and artificial intelligence applications are examples of this class.
\item `\emph{Storage-intensive}' applications require extensive storage resources. Decentralized storage and content distribution applications are examples of this class.
\item `\emph{Low cost}' applications are sensitive to the cost. Internet of things (IoT) and social media applications are examples of this class.
\item `\emph{Private}' applications require privacy guarantee.
\item `\emph{Management/Control}' class is used the control functions (e.g., resource allocation), management and governance functions of vDLT.
\item `\emph{Best effort}' describes a service in which the system does not provide any guarantee that service is delivered or that delivery meets any quality of service.
\item `\emph{Scavenger}' applications are those ones that are not desirable in the system.
\end{itemize}

\begin{table}
\label{table_CoS}
\centering
  \caption{Class of Service (CoS) in vDLT}
  \begin{tabular}{| c | c |}
\hline
  CoS Bit & Application
\\  \specialrule{1.3pt}{1pt}{1pt} 
1 1 1 & Fast confirmation \\ \hline 
1 1 0 & Computation-intensive \\ \hline 
1 0 1 & Storage-intensive \\ \hline 
1 0 0 & Low cost \\ \hline 
0 1 1 & Management/Control \\ \hline 
0 1 0 & Private \\ \hline 
0 0 1 & Best effort \\ \hline 
0 0 0 & Scavenger 

\\ \hline

\end{tabular}
\end{table}

\begin{figure*}[tp]
  \centering
	\vspace{-12cm}
	\includegraphics[width=1.4\textwidth]{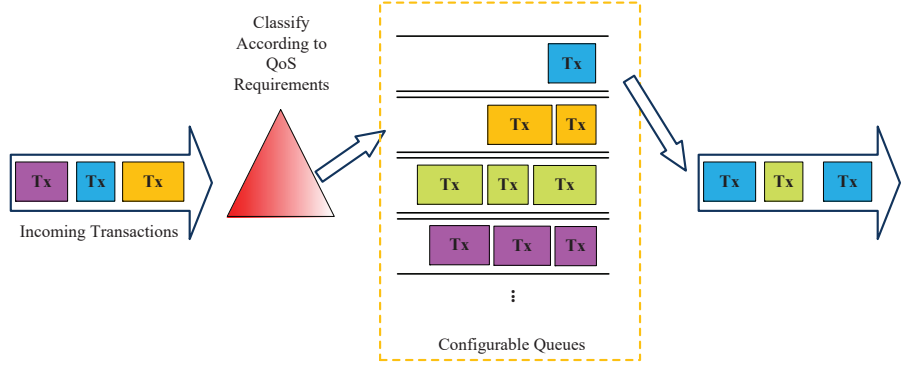}
	\vspace{-0.7cm}
		\caption{Classification and queuing of transactions at the control layer of vDLT.}
  \label{fig_queuing}
\end{figure*}

Due to the different QoS requirements, services and applications should be handled differently. Queuing mechanisms have been well studied and applied in the traditional Internet for QoS provisioning. In vDLT, we adopt advanced class-based weighted fair queuing (CBWFQ) \cite{FMS08}, which extends the standard WFQ functionality to provide support for defined  classes. With CBWFQ, transactions satisfying the match criteria for a class constitute the transactions for that class. A queue is reserved for each class, and transaction belonging to a class is directed to that class queue. After a class has been defined and its match criteria have been formulated, we can assign characteristics to the class according to the QoS requirements.

\subsection{Management/Control Nodes}
To guarantee decentralization, many DLT systems (e.g., Ethereum) require that every full node runs the smart contact, and checks that execution has gone correctly, which significantly affects the scalability of these DLT systems. Recently, various strategies have been proposed to address the scalability issue by letting less nodes execute the smart contract  (e.g., Lightning Network \cite{LightingNetwork}, Raiden Network \cite{RaidenNetwork}, Sharding and Plasma \cite{Plasma}, Cardano \cite{Cardano}, EOS \cite{EOS}, and Zilliqa \cite{Zilliqa}). 

From the system perspective, deciding which nodes to run the smart contract is one of the control functions in DLT systems, which is similar to deciding which routers to forward user traffic in the traditional Internet. From the evolution history of telephone networks, the traditional Internet and cellular networks, we learn that the control function should be decoupled from the execution of smart contracts in next generation DLT systems. In addition to the benefits of the decoupling seen in those systems, the decoupling in DLT systems can also enhance decentralization, because the management/control nodes do not produce user transaction blocks, which will be produced by the execution nodes in vDLT. This can help  address the centralization issues of some existing DLT systems (e.g., EOS) criticized by the community. This decoupling is similar to the separation of powers for the governance of a state, there the typical division is into three branches: a legislature, an executive, and a judiciary. 

\begin{figure}[tp]
  \centering
	\vspace{-12cm}
	\hspace{5cm}
  \includegraphics[width=1.45\textwidth]{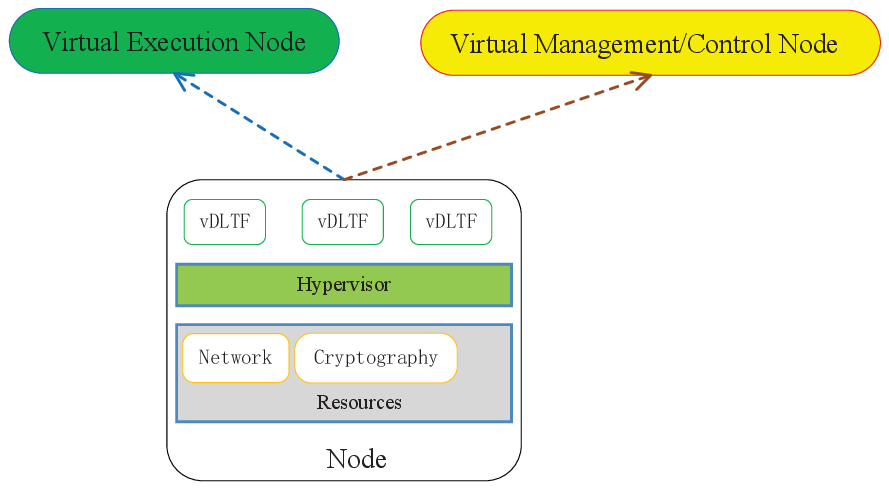}
	\vspace{-1.5cm}
		\caption{With virtualization, a node can be virtualized to a management/control node or a execution node.}
  \label{fig_vdlt_control_traffic}
\end{figure}

The decoupling of management/control from execution can be done via virtualization \cite{YLH18}. With virtualization, a node can be virtualized to a management/control node or execution node, as shown in Figure \ref{fig_vdlt_control_traffic}.

Management/Control nodes are responsible for the management/control functions of vDLT. They are required to have a stable performance, e.g., a dedicated IP address, running 24/7, high bandwidth, good hardware, etc. Management/Control nodes get paid of the reward on every management/control decision, which is distributed to management/control nodes one at a time. These management/control nodes do not produce user transaction blocks, which will be produced by the execution nodes in vDLT. The decoupling of management/control from execution can improve decentralization of vDLT.

\subsection{Delegated Proof-of-Stake (DPoS) and Quadratic Voting with Token Lock}
Management/Control nodes operate as part of the DPoS mechanism. Under DPoS, community members will vote on delegates to represent them on the system, and these delegates are charged with management/control functions. Unlike the plurality/majority voting systems used by most DPoS systems, quadratic voting \cite{LW17} with token lock \cite{QV16} is used in vDLT. Quadratic voting provides a better way to make collective decisions that avoids the tyranny of the majority. It allows people to express how strongly they feel about an issue rather than just whether they are in favor of it or opposed to it.  If a participant has a strong preference for or against a particular decision, additional weights can be allocated. However, the cost of additional weights increasingly becomes more expensive quadratically (e.g., 1 vote - \$1, 2 votes - \$4, 3 votes - \$9, 4 votes - \$16). In quadratic voting with token lock \cite{QV16}, which is used in vDLT, $N$ tokens let a participant make $N*k$ votes by locking up those tokens for a time period of $k^2$. It prevents a single group from quietly taking it over. It will take a group many cycles and a costly number of tokens to take control, likely alerting the rest of the blockchain users to the issue to take action.

\subsection{Consensus}
For the consensus among management/control nodes,  practical byzantine fault tolerance (PBFT) protocol is used with EC-Schnorr multi-signature \cite{STV16,KJG16}. With multi-signature, multiple signers aggregate their signatures into a single signature on a given message. A single public key that aggregates the keys of all the signers can be used to authenticate this singed message. Unlike the elliptic curve digital signature algorithm (ECDSA) used in Bitcoin and Ethereum, EC-Schnorr has been proven to be non-malleable \cite{Poe14}. The non-malleability property means that given a set of signatures generated on a message using a private key, it should be hard for an adversary to produce a new signature for the same message that is valid for the corresponding public key.

In addition, the use of EC-Schnorr multisignature lowers the normal case communication overhead from $O(n^2)$ in classical PBFT to $O(n)$ and reduces the signature size from $O(n)$ to $O(1)$, where $n$ is the size of the consensus group. Message authentication code (MAC) is used in classical PBFT for the authenticated messages exchanged among nodes. Since a share secret key is used in MAC, the classical PBFT has a communication overhead of $O(n^2)$, which make it impractical when the size of the consensus group is larger than 20. Inspired by ByzCoin \cite{KJG16} and Zilliqa \cite{Zilliqa}, MAC is replaced with digital signature to effectively reduce the communication overhead from $O(n^2)$ in classical PBFT to $O(n)$. Moreover, in classical EC-Schnorr multi-signature scheme, all the signers need to agree on signing a given message, and the signature is valid only if all the signers have signed the message. However, in iPBFT, only over $2n/3$ nodes are needed to sign the message. Therefore, a bitmap is used to indicate the nodes who participate in the signing process.

\subsection{Dynamic Resource Allocation}
Dynamic resource allocation is an important component in ``service-oriented" vDLT, will will satisfy the service-specific needs and at the same time optimize the use of scarce networking, storage, and computational resources. When making the decision on resource allocation and which nodes should execute the smart contact, the QoS class of the transaction and the state of the available execution nodes will be carefully considered. The algorithm is described as follows.

\begin{equation}
\label{eq_optim}
\begin{aligned}
\begin{array}{l}
{\cal P}1:\ \ \mathop {\max }\limits_{{\boldsymbol{\delta }},{\boldsymbol{\rho }},{\boldsymbol{s}}} \ \ \  Utility \\
\!s.t. \ C1: Decentralization \geq \gamma^{De}, \\
\ \ \  \ C2: ConfirmationLatency_{n} \leq \gamma_n^{CL}, \forall n, \\
\ \ \  \ C3: Throughput_{n} \geq \gamma_n^{Th}, \forall n, \\
\ \ \  \ C4: Cost_{n} \leq \gamma_n^{Co}, \forall n, \\
\ \ \  \ C5: Privacy_{n} = \gamma^{Pr}, \forall n, \\
\ \ \  \ \ldots \\
\end{array}
\end{aligned}
\end{equation}
In the above equation, $Utility$ is defined to measure the performance of the vDLT system. For example, $Utility$ can be defined as the overall throughput or the overall social welfare of the system. The system utility is optimized by controlling $\boldsymbol{\delta}, \boldsymbol{\rho}, \boldsymbol{s}$, where $\boldsymbol{\delta}$ represents the execution nodes that execute the smart contract, $\boldsymbol{\rho}$ represents the resource allocation in the management/control nodes, and $\boldsymbol{s}$ represents the resource allocation in the execution nodes. The first set of constraint $C1$ guarantees the degree of decentralization is satisfied. Constraint $C2$ ensures that the confirmation latency requirement of each transaction can be met. Here, $n$ is the transaction number in the system. Constraint $C3$ ensures that the throughput requirement can be met. Constraint $C4$ ensures that the cost requirement of each transaction can be met. Constraint $C5$ ensures that the privacy requirement of each transaction can be met. Here, $\gamma^{Pr} = \{0, 1\}$, where $0$ means no privacy is needed, and $0$ means privacy is needed. 

Please note that this formulation is general enough so that the system can easily evolve to incorporate other performance measures and constraints in the future.

\subsection{Deep Reinforcement Learning for Performance Optimization}
\begin{figure}[tp]
  \centering
	\vspace{-14.5cm}
	\includegraphics[width=1.5\textwidth]{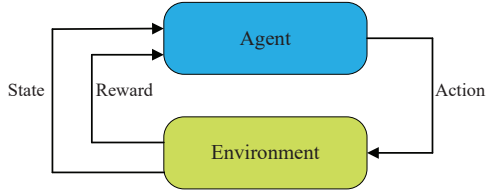}
	\vspace{-1cm}
		\caption{Deep reinforcement learning for performance optimization in vDLT.}
  \label{fig_DRL}
\end{figure}

In order to solve Eq. (\ref{eq_optim}), and optimize the performance of vDLT, we adopt a deep reinforcement learning approach in this work. Deep reinforcement learning is an advanced reinforcement learning algorithm that uses  deep neural networks  to approximate the value-action function \cite{mnih2015human}. Google Deepmind adopts this method on some games \cite{mnih2015human,SHMGS16}, and we have successfully used it for resource allocation problems in traditional networks \cite{HYZ17_ComMag,HLY18}. In deep reinforcement learning, an agent learns to take actions on the environment, and tries to obtain the most reward from the environment even though it faces with much uncertainty about the environment, as shown in Figure \ref{fig_DRL}. The agent has to make a tradeoff between the exploration and exploitation, and adjusts its actions based on the delayed rewards. Usually, a reinforcement learning problem can be described by using a Markov Decision Process (MDP). More advantageously, deep reinforcement learning can handle the complex situations in vDLT that the state, explicit transitional probability and immediate reward are not completely known, which makes this approach robust in practice. 

\subsection{Execution Nodes}

Execution nodes are responsible for the execution of smart contacts in vDLT. Due to the node heterogeneity, different execution nodes have different characteristics, and the state of the execution nodes should be reported to the management/control nodes. Some execution nodes have faster processors, some have higher trust values (i.e., more honest and trustworthy), while some have cheaper memory/storage. Therefore, in order to meet the different QoS requirements of different services and applications, different execution nodes should be dynamically selected to execute the smart contracts in a transaction, using the algorithm described in the above sections. For example, for applications requiring fast confirmation, more honest and less number of execution nodes should be selected. The effectiveness of a similar approach is shown in ThunderCore \cite{ThunderCore}. Once the execution nodes are selected, the transactions will be sent to these execution nodes. Similar to the management/control nodes, execution nodes use PBFT protocol with EC-Schnorr multi-signature for the consensus. For applications requiring high security, both  management/control nodes and execution nodes can be selected to execute the smart contracts to reach the consensus.

\subsection{Hierarchical Architecture}
\begin{figure}[tp]
  \centering
	\vspace{-10cm}
	\includegraphics[width=1.43\textwidth]{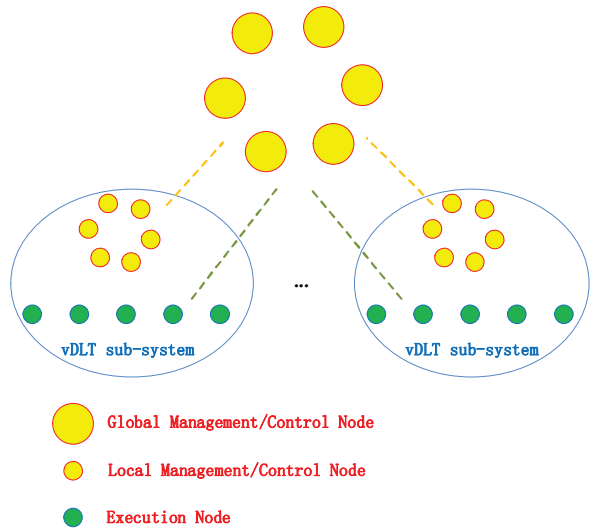}
	\vspace{-0.3cm}
		\caption{Hierarchical architecture of vDLT.}
  \label{fig_Hierarchical}
\end{figure}

Due to the the decoupling of management/control from execution and the centralization of the management/control logic in vDLT, scalability can become an issue. To address this issue and further improve the performance, a hierarchical architecture is used in vDLT, as shown in Figure \ref{fig_Hierarchical}. There are two layers for the management/control nodes. The bottom layer management/control nodes are responsible for the local services and applications that occur frequently near the local execution nodes. Multiple groups of local management/control nodes are deployed throughput the system; each group manage/control one of a handful of execution nodes. The top management/control nodes are responsible for the global services and applications that need a global view of the system. The hierarchical architecture can help achieve system scalability in vDLT. 

\subsection{Incentive Mechanisms}
The total number of vDLT tokens is 1 Billion with the potential of up to 4\% inflation per annum (depending on community votes). The new tokens from the inflation will be awarded to the management/control and execution nodes, which enables free transactions in vDLT. The number of tokens awarded is determined by the median of the desired pay by contributors. In addition, the tokens can be ``staked" to power the system's network, computation and storage capabilities. With the flexibility enabled by decoupling management/control and execution as well as dynamic resource allocation, the cost will be lower and the reward will be higher in vDLT compared to most existing DLT systems.

\subsection{Penalty Mechanisms}
The nodes in vDLT can together monitor the suspicious behaviors. A node will be explicitly penalized if misbehavior is found. If nodes put down
collatoral to participate, penalty can be implemented by taking away their collateral and rewarding the node who submits cryptographic evidence of misbehavior. Moreover, free-riding may occur in the system. For example, in an attempt to get free rewards, nodes who register to vote actually do not participate the voting process. Penalty should be implemented to  dis-incentivize this kind of behavior. This can be achieved by adjusting the reward mechanism to give more reward to those who have actively participated.

\subsection{Governance}
As with organisms, we believe that the most successful DLT systems will be those that can best adapt to their environments. Since DLT systems need to evolve to survive, initial design is important, but over a long enough timeline, the mechanisms for change in vDLT are important as well. As we have seen in the evolution history of telephone networks, the traditional Internet and cellular networks, the decoupling of management/control and execution facilitates independent evolution of the management/control plane and user plane functions. 

A robust on-chain mechanism is designed in vDLT that seamlessly amends the rules governing its protocol and rewards protocol development to enable vDLT a ``self-amending" system. Anyone can submit a change to the governance structure in the form of a code update. ``Multi-factorial consensus" \cite{NotesGovern} is used in vDLT, where different groups are polled, and the ultimate decision depends on the collective result of these polls together. The coordination includes the roadmap, core developers, token holders, users, and the established norms. Then, quadratic voting with token lock described above will occur. If it is passed, the update is first implemented on a test vDLT system. After a period of time on the test vDLT system, another vote takes place to confirm the change. If it is passed again, the change goes live on the main vDLT system.
 
\section{Conclusion}
The underlying distributed ledger technology (DLT) of crypto-currencies has great potential to create new foundations for our economic and social systems. However, most existing DLT systems do not distinguish the widely varying quality of service (QoS) requirements. In this work, we presented vDLT to address the challenges of the existing DLT systems. We first reviewed the development of telephone networks, the traditional Internet, and cellular networks, which had similar issues in the early stage of these systems. Inspired by the development of these systems, vDLT decouples management/control (e.g., governance, smart-contract-execution nodes selection, and resource allocation) and execution of smart contracts to support QoS provisioning, improve decentralization and facilitate evolution. vDLT represents a paradigm shift from the existing ``blockchain-oriented" DLT systems to next generation ``service-oriented" DLT systems. 

\balance
\bibliographystyle{IEEEtran}

\bibliography{references}

\begin{thebibliography}{10}
\providecommand{\url}[1]{#1}
\csname url@samestyle\endcsname
\providecommand{\newblock}{\relax}
\providecommand{\bibinfo}[2]{#2}
\providecommand{\BIBentrySTDinterwordspacing}{\spaceskip=0pt\relax}
\providecommand{\BIBentryALTinterwordstretchfactor}{4}
\providecommand{\BIBentryALTinterwordspacing}{\spaceskip=\fontdimen2\font plus
\BIBentryALTinterwordstretchfactor\fontdimen3\font minus
  \fontdimen4\font\relax}
\providecommand{\BIBforeignlanguage}[2]{{%
\expandafter\ifx\csname l@#1\endcsname\relax
\typeout{** WARNING: IEEEtran.bst: No hyphenation pattern has been}%
\typeout{** loaded for the language `#1'. Using the pattern for}%
\typeout{** the default language instead.}%
\else
\language=\csname l@#1\endcsname
\fi
#2}}
\providecommand{\BIBdecl}{\relax}
\BIBdecl

\bibitem{Bec18}
R.~Beck, ``Beyond bitcoin: The rise of blockchain world,'' \emph{Computer},
  vol.~51, no.~2, pp. 54--58, Feb. 2018.

\bibitem{IL17}
M.~Iansiti and K.~R. Lakhani, ``The truth about blockchain,'' \emph{Harvard
  Business Review}, Jan. 2017.

\bibitem{Bitcoin}
\BIBentryALTinterwordspacing
S.~Nakamoto, ``A peer-to-peer electronic cash system,'' Oct. 2018. [Online].
  Available: \url{https://bitcoin.org/bitcoin.pdf}
\BIBentrySTDinterwordspacing

\bibitem{LightingNetwork}
\BIBentryALTinterwordspacing
``Lighting network.'' [Online]. Available: \url{https://lightning.network/}
\BIBentrySTDinterwordspacing

\bibitem{RaidenNetwork}
\BIBentryALTinterwordspacing
``Raiden network.'' [Online]. Available: \url{https://raiden.network/}
\BIBentrySTDinterwordspacing

\bibitem{Plasma}
\BIBentryALTinterwordspacing
``Plasma.'' [Online]. Available: \url{http://plasma.io/}
\BIBentrySTDinterwordspacing

\bibitem{Cardano}
\BIBentryALTinterwordspacing
{Cardano}. [Online]. Available: \url{https://cardano.org}
\BIBentrySTDinterwordspacing

\bibitem{EOS}
\BIBentryALTinterwordspacing
``{EOS}.'' [Online]. Available: \url{https://eos.io}
\BIBentrySTDinterwordspacing

\bibitem{Zilliqa}
\BIBentryALTinterwordspacing
``{Zilliqa}.'' [Online]. Available: \url{https://zilliqa.com/}
\BIBentrySTDinterwordspacing

\bibitem{Jai90}
R.~Jain, ``Congestion control in computer networks: issues and trends,''
  \emph{IEEE Network}, vol.~4, no.~3, pp. 24--30, May 1990.

\bibitem{Jai92}
------, ``Myths about congestion management in high speed networks,''
  \emph{Internetworking: Res. and Exp.}, vol.~3, pp. 101--113, 1992.

\bibitem{UGM94}
B.~W. Unger, D.~J. Goetz, and S.~W. Maryka, ``Simulation of {SS7} common
  channel signaling,'' \emph{IEEE Comm. Mag.}, vol.~32, no.~3, pp. 52--62, Mar.
  1994.

\bibitem{BDO92}
M.~Bahl, J.~Daane, and R.~O'Grady, ``Evolving intelligent interexchange
  network-an ss7 perspective,'' \emph{Proceedings of the IEEE}, vol.~80, no.~4,
  pp. 637--643, April 1992.

\bibitem{IntServ}
R.~Braden, D.~Clark, and S.~Shenker, ``Integrated services in the {Internet}
  architecture: an overview,'' \emph{Internet RFC 1633}, June 1994.

\bibitem{DiffServ}
S.~Blake \emph{et~al.}, ``An architecture for differentiated services,''
  \emph{Internet RFC 2475}, Dec. 1998.

\bibitem{CO16}
B.~Chatras and F.~F. Ozog, ``Network functions virtualization: the portability
  challenge,'' \emph{IEEE Network}, vol.~30, no.~4, pp. 4--8, July 2016.

\bibitem{WRH15}
T.~Wood, K.~K. Ramakrishnan, J.~Hwang, G.~Liu, and W.~Zhang, ``Toward a
  software-based network: integrating software defined networking and network
  function virtualization,'' \emph{IEEE Network}, vol.~29, no.~3, pp. 36--41,
  May 2015.

\bibitem{NMN14}
B.~A.~A. Nunes, M.~Mendonca, X.~Nguyen, K.~Obraczka, and T.~Turletti, ``A
  survey of software-defined networking: Past, present, and future of
  programmable networks,'' \emph{IEEE Comm. Surveys Tutorials}, vol.~16, no.~3,
  pp. 1617--1634, Thirdquarter 2014.

\bibitem{Sal12}
V.~Salapura, ``Cloud computing: Virtualization and resiliency for data center
  computing,'' in \emph{Proc. IEEE 30th International Conference on Computer
  Design (ICCD)}, Sept. 2012, pp. 1--2.

\bibitem{MCD18}
R.~Morabito, V.~Cozzolino, A.~Y. Ding, N.~Beijar, and J.~Ott, ``Consolidate
  {IoT} edge computing with lightweight virtualization,'' \emph{IEEE Network},
  vol.~32, no.~1, pp. 102--111, Jan. 2018.

\bibitem{HGJ15}
B.~Han, V.~Gopalakrishnan, L.~Ji, and S.~Lee, ``Network function
  virtualization: Challenges and opportunities for innovations,'' \emph{IEEE
  Comm. Mag.}, vol.~53, no.~2, pp. 90--97, Feb 2015.

\bibitem{G-Lab}
\BIBentryALTinterwordspacing
G-lab. [Online]. Available: \url{http://www.german-lab.de/}
\BIBentrySTDinterwordspacing

\bibitem{achemlal2009virtualisation}
M.~Achemlal, T.~Almeida, and etc., ``D-3.2.0 virtualisation approach:
  Concept,'' The FP7 4WARD Project, Tech. Rep., 2009.

\bibitem{LY15}
C.~Liang and F.~R. Yu, ``Wireless network virtualization: A survey, some
  research issues and challenges,'' \emph{IEEE Commun. Surveys Tutorials},
  vol.~17, no.~1, pp. 358--380, Firstquarter 2015.

\bibitem{CUPS}
\BIBentryALTinterwordspacing
{3GPP}, ``{Control and User Plane Separation of EPC nodes (CUPS)}.'' [Online].
  Available: \url{http://www.3gpp.org/cups}
\BIBentrySTDinterwordspacing

\bibitem{LW17}
\BIBentryALTinterwordspacing
S.~Lalley and E.~G. Weyl, ``Quadratic voting: How mechanism design can
  radicalize democracy,'' \emph{American Economic Association Papers and
  Proceedings}, vol.~1, no.~1, December 2017. [Online]. Available:
  \url{https://ssrn.com/abstract=2003531 or
  http://dx.doi.org/10.2139/ssrn.2003531}
\BIBentrySTDinterwordspacing

\bibitem{QV16}
\BIBentryALTinterwordspacing
V.~Buterin, ``{On Coin-lock voting, {Futarchy} and Optimal Decentralized
  Governance}.'' [Online]. Available:
  \url{https://www.reddit.com/r/ethereum/comments/4rtpmm}
\BIBentrySTDinterwordspacing

\bibitem{STV16}
E.~Syta, I.~Tamas, D.~Visher, D.~I. Wolinsky, P.~Jovanovic, L.~Gasser,
  N.~Gailly, I.~Khoffi, and B.~Ford, ``Keeping authorities ``honest or bust"
  with decentralized witness cosigning,'' in \emph{Proc. IEEE Symposium on
  Security and Privacy}, May 2016, p. 526–545.

\bibitem{KJG16}
E.~Kokoris-Kogias, P.~Jovanovic, N.~Gailly, I.~Khoffi, L.~Gasser, and B.~Ford,
  ``Enhancing bitcoin security and performance with strong consistency via
  collective signing,'' in \emph{Proc. 25th USENIX Security Symposium, USENIX
  Security 16}, Aug. 2016, p. 279–296.

\bibitem{FMS08}
M.~J. Fischer, D.~M.~B. Masi, and J.~F. Shortle, ``Simulating the performance
  of a class-based weighted fair queueing system,'' in \emph{Proc. 2008 Winter
  Simulation Conference}, Dec. 2008, pp. 2901--2908.

\bibitem{YLH18}
F.~R. Yu, J.~Liu, Y.~He, P.~Si, and Y.~Zhang, ``Virtualization for distributed
  ledger technology ({vDLT}),'' \emph{IEEE Access}, vol.~6, pp.
  25\,019--25\,028, 2018.

\bibitem{Poe14}
\BIBentryALTinterwordspacing
A.~Poelstra, ``Schnorr signatures are non-malleable in the random oracle
  model,'' Feb. 2014. [Online]. Available:
  \url{https://download.wpsoftware.net/bitcoin/wizardry/schnorr-mall.pdf}
\BIBentrySTDinterwordspacing

\bibitem{mnih2015human}
V.~Mnih, K.~Kavukcuoglu, D.~Silver, A.~A. Rusu, J.~Veness, M.~G. Bellemare,
  A.~Graves, M.~Riedmiller, A.~K. Fidjeland, G.~Ostrovski \emph{et~al.},
  ``Human-level control through deep reinforcement learning,'' \emph{Nature},
  vol. 518, no. 7540, pp. 529--533, Feb. 2015.

\bibitem{SHMGS16}
D.~Silver, A.~Huang, C.~J. Maddison, A.~Guez, L.~Sifre, G.~van~den Driessche,
  J.~Schrittwieser, I.~Antonoglou, V.~Panneershelvam, M.~Lanctot \emph{et~al.},
  ``Mastering the game of go with deep neural networks and tree search,''
  \emph{Nature}, vol. 529, no. 7587, pp. 484--489, Jan. 2016.

\bibitem{HYZ17_ComMag}
Y.~He, F.~R. Yu, N.~Zhao, V.~C.~M. Leung, and H.~Yin, ``Software-defined
  networks with mobile edge computing and caching for smart cities: A big data
  deep reinforcement learning approach,'' \emph{IEEE Comm. Mag.}, vol.~55,
  no.~12, pp. 31--37, Dec. 2017.

\bibitem{HLY18}
Y.~He, C.~Liang, F.~R. Yu, and Z.~Han, ``Trust-based social networks with
  computing, caching and communications: A deep reinforcement learning
  approach,'' \emph{IEEE Trans. Network Science and Eng.}, May 2018.

\bibitem{ThunderCore}
\BIBentryALTinterwordspacing
``{ThunderCore}.'' [Online]. Available: \url{https://www.thundercore.com/}
\BIBentrySTDinterwordspacing

\bibitem{NotesGovern}
\BIBentryALTinterwordspacing
V.~Buterin, ``{Notes on Blockchain Governance}.'' [Online]. Available:
  \url{https://vitalik.ca/general/2017/12/17/voting.html}
\BIBentrySTDinterwordspacing

\end{thebibliography}

\end{document}